# Ethnographie de la structuration d'un corpus collectif de messages de soutien social en ligne.

Goritsa Ninova, Hassan Atifi
ICD/Tech-CICO, Université de Technologie de Troyes
12 rue Marie Curie, BP. 2060, 10010 Troyes Cedex
{goritsa.ninova, hassan.atifi }@**utt.fr**

RÉSUMÉ. Dans cet article, nous proposons une étude de l'élaboration progressive de la structure d'un corpus d'«échanges de soutien social» recueilli à partir des forums de discussion. Il s'agit des premiers résultats d'une description ethnographique des pratiques de constitution et d'analyse d'un corpus, situées dans le cadre d'un projet interdisciplinaire. Dans un premier temps, nous décrivons le dossier numérique « personnel » d'un chercheur, relatif à son l'analyse d'un corpus d'interaction de soutien social en ligne. Ensuite, nous examinons la constitution d'un corpus partagé, résultat de la mise en commun de corpus individuels, de petites tailles, produits par les chercheurs participants au projet, chaque corpus individuel reflétant les perspectives de recherche de son « producteur ».

ABSTRACT. In this paper, we propose a study of progressive development of the structure of corpus collected starting from the discussion forums. These are the first results from an ethnographic description of constitution and analysis practices of a corpus located within an interdisciplinary project. At first, we describe the personal digital documents folder of a researcher, relative to his corpus analysis of interactions of social support online. Next, we examine the constitution of a shared corpus, result of the pooling of individual corpora, small size, produced by researchers participating in the project, each individual corpus reflecting his producer's perspectives of research.

MOTS-CLÉS : anthropologies des connaissances, éthographies des pratiques de constitution et d'analyse de corpus, corpus d'interactions de soutien social en ligne.

## Introduction

Dans cet article, nous proposons une étude de l'élaboration progressive de la structure d'un corpus collectif de «messages de soutien social» recueilli à partir des forums de discussion. Il s'agit des premiers résultats d'une description ethnographique des pratiques collectives de constitution et d'analyse d'un corpus, situées dans le cadre d'un projet interdisciplinaire qui réunit des chercheurs en sciences humaines et sociales et des chercheurs en informatique. La structure du corpus collectif est étudiée à travers deux études de cas. Dans un premier temps, nous décrivons le dossier numérique « personnel » d'un chercheur, relatif à son l'analyse d'un corpus d'interaction de soutien social en ligne. Ensuite, nous examinons la constitution d'un corpus partagé, résultat de la mise en commun de corpus individuels, de petites tailles, produits par les chercheurs participants au projet, chaque corpus individuel reflétant les perspectives de recherche de son « producteur ». L'objectif est d'appréhender ce mode particulier de production des connaissances dont le corpus est à la fois le résultat des pratiques concrètes de constitution et l'objet d'un travail d'analyse. Les résultats de cette étude peuvent apporter des préconisations quant à l'instrumentation des pratiques collectives de constitution et d'analyse de corpus pour les études de l'interaction. Ces préconisations peuvent être exprimées en termes de choix des techniques et des outils documentaires les mieux appropriés par rapport aux pratiques observées

## 2. Qu'entend-on par corpus ?

Contrairement à une archive, le corpus présente les caractéristiques spécifiques suivantes : i) il est construit pour les besoins d'une problématique de recherche [HAB 00] [CON 05] ; ii) il constitue l'objet d'une analyse dans le cadre de la recherche pour laquelle il a été construit [CHA 09]; iii) Il offre à l'analyste de multiples vues sur son objet de recherche. Par le jeu de la mise en parallèle de divers niveaux de description, le chercheur prend distance avec son matériau de recherche et atteint une certaine généralisation des résultats [MON 05] [TRAV 99], [RAS 05].

Ces trois caractéristiques sont discutées dans des écrits d'ordre épistémologique et méthodologique en termes de clôture du corpus et de prise en compte des différents contextes lors de l'analyse. La problématique de clôture et de contexte peut être traitée également du point de vue de l'emboitement de sous-corpus ou bien de déconstruction/construction moyennant l'application de critères de contraste [RAS 05], [CHA 09]. Ceci souligne l'importance pragmatique de la structuration du corpus (celle qui explicite les liens entre les différentes parties constitutives du corpus) dans la construction du sens à partir du corpus et le fait qu'elle est créée progressivement dans l'activité du chercheur.

Notre problématique se décline en quelques questions de recherche : La dépendance entre le corpus et la problématique restreint-elle la possibilité de mutualisation des corpus ? Il y a-t-il des inscriptions directement liées à l'analyse ? Dans une optique de partage des connaissances, quels sont les éléments qui permettent à d'autres chercheurs de comprendre la problématique inhérente au corpus dans l'objectif de contraster leurs corpus?

## 3. Cadre de l'analyse

Globalement, notre approche s'apparente à de l'observation ethnographique de l'activité scientifique [LAT 87] et de conception [VIN 99]. Comme proposé par ce dernier, l'objectif est de comprendre l'activité, d'exploiter ensuite les résultats obtenus pour la soutenir et l'instrumenter. Nous focalisons notre attention sur le corpus - l'objet central, à la fois produit et support, de l'activité de recherche. Dans la démarche suivie nous étudions la fabrication progressive de la structure du corpus collectif à travers les contributions individuelles et collectives des participants au projet. Pour ce faire nous examinons de près la façon dont celle-ci dépend non seulement de la problématique commune de recherche, mais également de la spécificité du matériau recueilli et des thèmes de recherches propres à chaque chercheur.

Notre position est double : a) de point de vue de la publication d'un corpus, l'interdépendance entre le corpus et la problématique nécessite que le corpus soit présenté aux autres chercheurs avec les traces de son analyse (par exemple avec la grille d'analyse). Par conséquent, la grille d'analyse fait partie de la structure générale du corpus et permet des vue sur le corpus selon des critères de contraste définis lors de l'analyse. b) la construction d'un corpus partagé par la mutualisation des corpus individuels, induit la réalisation d'opérations de mise en contraste sur les corpus constituant. Cette réalisation nécessité la mobilisation d'un espace de publication [IAC 06] pour assurer la construction (la négociation) d'un savoir de référence par rapport à l'expression de différent points de vue.

Pour étayer notre propos, présentons deux études de cas : l'organisation d'un dossier « personnel » et le mode de construction d'un corpus collectif. Dans le premier cas, nous décrivons : i) comment les documents sont rangés ? ii) comment sont-ils structurés ? iii) quels sont les indices qui lient le corpus à la problématique ? Dans le second, nous partons des traces de l'analyse qui subsistent pour voir comment le corpus collectif peut être fabriqué ?

# 4. Etudes de cas

## *3.1. Le dossier de travail « Corpus Lille »*

Le premier cas est consacré au dossier numérique « Corpus Lille1» de M., linguiste conversationnaliste. Ce dossier contient le corpus et les résultats d'analyse relatifs à une première étude d'extraits du forum Doctissimo (http://forum.doctissimo.fr/). Il s'agit d'échanges de soutien social. Cette étude est conduite en collaboration avec N., chercheure en psychologie cognitive. L'objectif des chercheurs est double. Tout d'abord, ils souhaitent vérifier en corpus les catégories psycho-sociales de soutien social. Ensuite, ils visent la production d'un script type pour chaque catégorie de soutien social étudié.

### 4.1.1. Vue générale sur le dossier « Corpus Lille »

Le dossier contient quatre dossiers aux noms évocateurs: « Analyse brouillon », « Analyse corpus », « Brouillon » , « Corpus » et des documents non rangés dans des dossiers, nommés respectivement « Notes », « Présentation Lille» et « Analyse requêtes + corpus ».

Dans les descriptions qui suivent, nous étudions des traces matérielles qui traduisent la dépendance du corpus à l'égard des tâches que le chercheur lui aurait appliqué en fonction de la problématique: sélection, découpage, codage, description, rangement, interrogation, etc.

### 4.1.1.1. Le sous-dossier « Corpus »

Il est indéniable que le corpus est un construit, mais comment cette construction se réalise-t-elle dans la pratique ? L'examen de ce dossier, nous permettra de répondre à cette question.

Le dossier « Corpus » contient 10 documents aux formats HTML et autant de documents au format Word, copies des mêmes Fils de discussion extraits de divers forums de Doctissimo. Les fichiers HTML contiennent la copie de la totalité des messages (au moment de l'enregistrement), tandis que les fichiers Word ne contiennent que les 10 premiers messages de chaque Fils choisi. C'est le premier « emboîtement » que nous observons. Par rapport aux noms des fichiers, les fichiers HTML, reprennent dans leur nom, donné par défaut par le navigateur, la structure des forums dans le site2, tandis que les noms des fichiers Word, choisis par le chercheur, ne reprennent que le nom du Fil, tiré à partir du titre du premier message. Ce nom est souvent tronqué. Par exemple, le nom du Fil « Connaissez vous le spasmine est il efficace » dans la version Word devient « spasmine ».

L'examen du dossier « Analyse corpus », nous permettra de comprendre le sens des emboitements de corpus construits lors de l'analyse, et leur dépendance de la problématique de recherche. Il nous renseigne également au sujet de la façon dont le chercheur exploite la structure documentaire du site Doctissimo3.

---

[1] C'est le nom du dossier principal. Le nom de la ville est donné par les chercheurs en référence à la conférence où les résultats sont présentés. Nous l'avons changé par respect des consignes d'anonymisation.

[2] Par exemple dans le nom du fichier « Pour Virginie - Affections neurologiques (Alzheimer, brParkinson, SEP) - FORUM Santé . html». « Pour Virginie » est le nom du Fil, il est séparé par un tiré du nom d'un sous-thème, ici Alzheimer, brParkinson, SEP. Ce dernier, pour ça part, est séparé du nom du forum, ici le « Forum Santé ». Ces informations sont reprises dans les documents Word, où le chercheur fait systématiquement une copie de la représentation graphique de l'emplacement du fil dans les archives du site.

### 4.1.1.2. Le sous-dossier « Analyse Corpus »

Le dossier « Analyse corpus » contient des documents Word nommés Fil1, Fil2, etc., ainsi que les trois documents: « Messages initiatifs », « Messages évaluatifs » et « Grille d'analyse messages de soutien ». Le document FIL1 est composé de 3 parties distinctes, une relative à la structure du Fils, et deux autres comportant des tableaux. Le premier tableau porte le titre « Requête et autres messages de celui qui produit la requête », le tableau suivant –« Réactions à la requête ». Les lignes du premier tableau comportent des catégories provenant d'une grille d'analyse relative au message de demande de soutien. Les lignes du second tableau, contiennent les catégories de la grille « grille message de soutien », relative à l'analyse des messages du soutien. En colonnes des deux tableaux sont inscrits les messages (ex. message1, message2). Les colonnes contiennent d'autres informations relatives au message analysé, par exemple le pseudonyme de celui qui poste le message (ex. nizou38), ainsi que des catégories relative au type du message en fonction de ça position dans le Fil de discussion (ex. requête, réaction au soutien). Dans les cases des tableaux sont copiés des fragments des messages qui correspondent à la catégorie concernée, ou des commentaires paraphrasant le contenu du message.

Les documents « Messages initiatifs » et « Messages évaluatifs » comportent des tableaux de synthèse. Par exemple, le tableau « Messages initiatifs » contient la même grille d'analyse que le tableau dans le document Fil1, par contre en colonne on trouve tous les messages de requêtes de tous les documents d'analyse nommés Fil (numéro-fil). Une information supplémentaire, indiquant le numéro du Fil, figure en plus du type de message et le pseudonyme du poseur.

### 4.1.4 Les multiples structures du corpus « Corpus Lille »

Qu'apprend-on au sujet la composition du corpus ? Tout d'abord, le corpus « Corpus Lille » est un corpus de Fils de discussion, car l'analyse successive des Fils dans leur « déroulement » semble être très importante pour le chercheur. Ensuite les messages reçoivent d'une part des catégories « requête », « réponse à la requête » et « réponse au soutien » et de l'autre part des catégories « Demande de soutien émotionnel », « Demande de partage d'expériences », « Demande de soutien tangible » et « Demande d'évaluation et de partage d'expériences ». Les fragments des messages reçoivent les catégories de la grille d'analyse (ex. Ouverture, Description de l'activité dans le forum, Formulation de l'action faite pour l'autre). Par ailleurs, la composition des tableaux synthétiques semble indiquer que les Fils de discussion sont rangés eux-aussi dans des catégories relatives au type de la demande. Nous notons que durant l'analyse le chercheur résonne effectivement en termes de sous-corpus. Ces derniers peuvent être constitués de messages, mais également de fragments de messages.

Pourtant, une partie du raisonnement du chercheur échappe à l'observation. Contrairement aux catégories précédentes, l'unité d'analyse de base ne sont visibles ni à la lecture de son dossier, ni à partir du support de présentation orale. Il s'agit du rapprochement fait entre un forum et une conversation. Il s'ensuit que les fils de discussion sont associés à des séquences d'échange dans une conversation et les messages aux interventions. Ces catégories ne sont pas explicitées. Les raisons peuvent être de deux natures : 1) Ces catégories sont largement partagés par la communauté scientifique qui analyse le langage en contexte et par conséquent elles n'ont pas besoin d'être reprises. 2) En ce qui concerne le rapprochement entre les catégories analytiques relatives à l'analyse d'une conversation et les forums de discussion, celui-ci est discutée dans des articles à caractère méthodologiques écrit par le chercheur et fait partis de l'histoire conversationnelle des échanges dans ce champs de recherche.

---
[3] Un forum de discussion peut être vue comme une conversation à plusieurs. Cependant, il s'agit d'un dispositif sociotechnique qui propose un archivage des messages au sein d'un Fil de discussion. Ce-dernier est rangé avec d'autres Fils de discussions dans des dossiers selon différents thématiques.

On peut conclure que la structure résultante du corpus « Corpus Lille » regroupe les structures relatives aux différents niveaux d'analyse, actualisés par le chercheur dans l'application de critères contrastés. Par exemple, à partir du support de la présentation orale de l'étude, nous observons des constructions qui laissent entrevoir le mode d'interrogation du corpus par le chercheur. Ce dernier souhaite savoir quelles sont les micro-activités (catégories associés à des fragments) qui sont typiques aux demandes de soutiens émotionnels (catégories associés aux messages), ou bien comment habituellement se réalise une interaction de soutien social ? (Croisement de catégories associé aux fils avec des catégories associés aux fragments).

## *3.2. La construction du corpus collectif*

Le corpus « Corpus Lille » entre ensuite dans la composition d'un corpus collectif, créé dans le cadre du projet MISS[4]. Il s'agit d'un projet réunissant sept chercheurs dont deux linguistes, un sociologue, une psychologue et trois informaticiens. Ce projet se caractérise par l'interdisciplinarité et les relations horizontales entre les participants. Les chercheurs qui participent à la constitution et l'analyse du corpus (les deux linguistes, le sociologue et la psychologue) entretiennent une relation d'interdépendance entre le thème du projet (comment soutenir le soutien social en ligne) [LEW 08] et les propres travaux de chacun. Ce qui se concrétise par la rédaction collective d'articles de recherche en parallèle avec un travail de validation des propres travaux de chacun, en fonction des règles de validité de sa discipline.

Nous passons en revue l'essentiel des spécificités de deux autres corpus de forum de discussion qui rentrent dans la composition du corpus partagé pour poser ensuite la question du mode de construction du corpus résultat – le corpus « MISS »

### 3.2.1. Les autres corpus du projet MISS

#### 3.2.1.1. Le corpus « Nouveau corpus »

Il s'agit d'un corpus construit par N., la psychologue, pour une étude sur les conditions de réalisation du soutien social en ligne, publiée dans une revue de psychologie. Dans cette étude, elle reprend la grille d'analyse, issue de la collaboration avec le linguiste, en construisant un nouveaux corpus à partir du même terrain (le site Doctissimo). Nous nous intéressons à la répercussion de la problématique sur la structure du nouveaux corpus.

---

[4] Le projet « Modèle de l'Internet pour le Soutien Social » (MISS) est un projet stratégique financé par l'établissement pour une durée de 3ans. L'objectif du projet est « de proposer un modèle du soutien social sur Internet, basé sur l'analyse d'interactions de soutien social en ligne et sur les diverses théories en psychologie et sociologie, afin de définir des fonctionnalités spécifiques adaptées à cette activité qui pourront être proposées à des organisations souhaitant offrir un service de soutien en ligne ».

Ainsi, comparée au corpus « Corpus Lille », constitué principalement de Fils de discussion représentant des séquences de conversation sur des thématiques variées, le corpus « Nouveau corpus » est constitué de message avec ou sans réponse, portant sur deux problèmes de santé : le cancer et la migraine. Par rapport à la grille initiale, nous observons moins de catégories. En revanche la grille résultante est plus structurée. La chercheure a procédé à des regroupements de catégories sous des catégories plus génériques. On observe également une reformulation de quelques catégories (exemple : Description de l'activité dans le forum, devient *Description de la connaissance et de l'usage du forum* ; Etat psychologique devient *Description de l'expérience émotionnelle accompagnant le problème*). Ces modifications visent la fabrication d'une grille plus adaptée aux objectifs de l'étude, l'interprétation des catégories par rapport aux facteurs du soutien social entre proches, ainsi que la possibilité d'appliquer un traitement statistique.

### 3.2.1.2. Les corpus « Ankara »

Le corpus « Ankara »[5] est construit par H, le deuxième linguiste de l'équipe, en vue d'une étude sur la spécificité de l'entraide numérique dans un forum diasporique. Cette étude, présentée à une conférence en sociologie de la communication, est conduite en collaboration avec le sociologue. Le corpus « Ankara » diverge des deux premiers corpus par la problématique et le terrain. Tout d'abord, contrairement aux deux études précédentes, les chercheurs font recours à la notion d'entraide numérique et non à celle de soutien social en ligne, le soutien social représentant une des réalisations possibles de l'entraide. Par ailleurs, ce corpus est prélevé à partir du forum diasporique Bladi (http://www.bladi.net/). Ce fait a une incidence non seulement par rapport à l'organisation des sous-corpus des chercheurs, mais également par rapport aux catégories relatives au type de soutien social, car la grille d'analyse reflète essentiellement ce que le chercheur appelle « entraide technique ». C'est le type d'aide qui est le plus souvent rencontré dans ce forum. Par exemple, la grille comporte dix types de demande d'entraide en prenant en compte la teneur des propos échangés : informationnelle, administrative, technique, linguistique, scolaire, professionnelle, religieuse, entraide psychologique, entraide politique ou militante, autres.

La comparaison avec les deux précédant corpus confirme :i) que le changement de problématique induit des modifications dans la structure du corpus et dans la grille d'analyse ; ii) que la structure du corpus et la grille d'analyse reflètent la spécificité du terrain. Toutefois, les trois corpus demeurent comparables, du fait de leur matérialité commune et l'approche partagé. Il s'agit de corpus de forum de discussion et des grilles partagent les mêmes unités d'analyse, celles qui décrivent les interactions verbales en unité de rangs différents : séquence, échange, intervention, etc. Ces unités sont liées de façon similaire aux diverses unités documentaires dans les corpus : les documents entiers, les segments et les fragments des documents. La comparaison fine des grilles d'analyse pointe vers des catégories d'analyse communes aux trois grilles (par exemple, ouverture, clôture, présentation du soi, etc.). Ces similitudes entres les corpus s'expliquent par les cadres théoriques partagés par les deux linguistes, ayant reçu la même formation.

Il reste à savoir si le corpus construit par la mutualisation de ces corpus représente un unique fond documentaire. Nous proposons de suivre la démarche propre aux chercheurs.

---

[5] Comme pour le corpus « Corpus Lille », nous reprenons le nom du dossier principal. Le nom de la ville est changé.

### 3.2.2. Le corpus collectif du projet Miss

Ainsi, le corpus collectif du projet MISS est construit par la mutualisation de plusieurs corpus individuels. Il résulte d'opérations successives de mise en correspondance de ces corpus-constituants[6]. Ces opérations sont réalisées à partir d'un travail d'harmonisation des structures initiales et notamment les grilles d'analyses relatives à chaque corpus. Nous appelons ce type de corpus « corpus partagé » (Ninova et Atifi, 2009). La mise en contraste, induit de nouveaux ajouts dans la structure du corpus. Par exemple, opposition des sous-corpus issus des forums de discussions différents, la comparaison des échanges de soutien social en ligne (les corpus de forum de discussion) avec des échanges en pressenties (des corpus d'observation de focus groupe). Notons que toutes les descriptions appliquées sur les corpus sont porteuses de sens pour les chercheurs.

A l'issus de ces descriptions, nous proposons d'expliciter la structure du corpus en distinguant trois parties constitutives: i) une structure endogène, propre à la structuration en section des documents d'origine (pour les forums de discussion, les extraits sont organisés en fils de discussion, chacun composé de messages) ; ii) une structure analytique , induite par l'analyse en profondeur et représentée par la grille d'analyse (par exemple, analyse de chaque message décrivant les micro-actes[7] du langage) ; iii) et enfin une structure documentaire, résultant d'une part, de la description des documents au sein du corpus et de l'autre, de la description du corpus entier au sein du corpus collectif du projet (par exemple, pour le projet MISS, il s'agit d'ajout des noms de forum, du type de l'échange (en présence versus en ligne etc.).

## 4. Discussion et Conclusion

Dans cet article, nous mettons en exergue des pratiques de constitution et d'analyse de corpus qui divergent du modèle du partage de corpus basé sur un hypothétique consensus. Suivre ce modèle conduirait les chercheurs, vers la normalisation des corpus constituants et la standardisation de la grille d'analyse. Or, les structures des corpus exprimant le point de vue du chercheur ne sont pas connues en avance. Elles peuvent faire objet d'une harmonisation en fonction de l'activité à postériori. Par conséquent, par rapport à l'organisation des connaissances se présente l'opportunité d'adopter un modèle de descriptions apte à représenter des "connaissances" d'un domaine selon différents points de vue [IAC 06], [Mas 08].

L'harmonisation des points de vue, à partir de la mise en contraste des éléments des corpus constituants, pose de nouveaux défis : Quelle instance décidera de la mise en commun des catégories et de la constitution des sous-corpus ? Quelle infrastructure pour négocier ce nouvel objet documentaire ? Qui réalise les opérations de mise en correspondance entre les catégories ? Qui est le destinataire final du corpus : les chercheurs qui manipulent la matérialité langagière et textuelle du corpus, l'équipe interdisciplinaire en vue d'une meilleure intercompréhension ; un champ disciplinaire en vue de la généralisation des résultats d'études à caractères local [MON 05]. Ces questionnements ouvrent de nouvelles pistes de recherche.

Pour conclure, à partir de ces deux études de cas, nous mettons en avant les spécificités de la structure du corpus collectif par rapport à un projet de recherche interdisciplinaire. Nous démontrons que les pratiques de construction et d'analyse de corpus observées sont assimilables à des pratiques documentaires dynamiques. Pour ce faire nous nous référons à une approche à la description structurée d'objets documentaires qui intègre les pratiques d'annotation (description d'un fragment parmi un document « sur mesure »), de rédaction (description d'une section parmi un document) et d'indexation (description d'un document parmi une collection) [BEN 04].

---

[6] Ces opérations sont repérables dans les articles du projet MISS écrits à plusieurs voix. Elles nous indiquent les critères de contraste qui peuvent être appliqué dans le but de fabriquer un corpus partagé.
[7] Les micro-activités constitutives des interactions de soutien dans le forum.

## Bibliographie